%% file: paper.tex
\documentclass[sigconf,nonacm]{acmart}

\usepackage{booktabs}
\usepackage{multirow}
\usepackage{xspace}
\usepackage{algorithm}
\usepackage{algpseudocode}
\usepackage{tikz}
\newcommand{\lastedit}{}
\usetikzlibrary{arrows.meta,decorations.pathreplacing}

\newcommand{\abc}{\textsf{abc}\xspace}
\newcommand{\abcpacked}{\textsf{abc-packed}\xspace}
\newcommand{\abcbmms}{\textsf{abc-bmms}\xspace}

\newcommand{\abcbmmspacked}{\textsf{abc-bmms-packed}\xspace}
\newcommand{\bmp}{\textsf{BMP}\xspace}

\newcommand{\spidx}{\textsf{SP}\xspace}
\newcommand{\lsp}{\textsf{LSP}\xspace}

\newcommand{\ioqp}{\textsf{IOQP}\xspace}

\newcommand{\rise}{\textsf{RISE}\xspace}
\newcommand{\qdrant}{\textsf{Qdrant}\xspace}
\newcommand{\seismic}{\textsf{Seismic}\xspace}

\newcommand{\kannolo}{\textsf{kANNolo}\xspace}
\newcommand{\pisa}{\textsf{PISA}\xspace}
\newcommand{\lucene}{\textsf{Lucene}\xspace}

\setcopyright{none}
\renewcommand\footnotetextcopyrightpermission[1]{}
\acmDOI{}
\acmISBN{}

\begin{document}

\title{The \abc index: Exact Retrieval for Learned Sparse Representations}

\author{Rossano Venturini}
\affiliation{%
  \institution{University of Pisa}
  \city{Pisa}
  \country{Italy}}
\email{rossano.venturini@unipi.it}

\begin{abstract}
In learned sparse retrieval (LSR), queries have tens of terms with learned
weights, on which classic dynamic pruning is believed to fail, and
approximate indexes such as \seismic and HNSW are considered faster than
exact ones. We challenge this view with \abc, an exact index built only from standard IR techniques: term-at-a-time (TAAT)
evaluation, Block-Max MaxScore, binary packing and doc-id reordering. They
are combined in a new design and engineered for modern CPUs. On MS MARCO
encoded with four LSR models, the \abc index is the fastest
exact index at every $k$: 6.5--84$\times$ faster than the exact search of
\qdrant, 13--55$\times$ faster than that of \lucene, and 2.9--18$\times$
faster than \bmp with an index 5--7$\times$ smaller. Against
approximate indexes, it takes 1.5--3.6$\times$ less space, and its latency
grows much more slowly with $k$: they are up to 4.3$\times$ faster at
$k=10$, but at $k=1000$ it overtakes them, by up to 7.4$\times$.
Exactness comes with an extra benefit: there are no parameters to tune.
\end{abstract}

\ccsdesc[500]{Information systems~Search engine indexing}
\ccsdesc[500]{Information systems~Retrieval efficiency}

\keywords{learned sparse retrieval, exact top-\texorpdfstring{$k$}{k} retrieval, term-at-a-time
query processing, dynamic pruning, inverted index compression}

\maketitle

\section{Introduction}
\label{sec:intro}

Learned sparse retrieval (LSR) models such as
SPLADE~\cite{formal2021splade,formal2022distillation} encode queries and
documents as sparse vectors over the vocabulary, and retrieve the $k$
documents with the largest inner product with the query. They keep the
interpretability and the infrastructure of the inverted index, but they
break the assumptions on which decades of query processing were built:
queries have tens of terms instead of a handful, and their weights are
learned rather than derived from term statistics, the ``wacky weights'' of
Mackenzie et al.~\cite{mackenzie2021wacky}. On these queries the classic
dynamic pruning algorithms, such as WAND and Block-Max WAND, can be slower
than scoring every posting~\cite{mackenzie2021wacky,mallia2024bmp}.

Two families of indexes have answered this challenge. Exact indexes, such
as \bmp~\cite{mallia2024bmp}, \spidx~\cite{carlson2025sp} and
\lsp~\cite{carlson2026lsp}, reorganize the collection into blocks of
documents with precomputed score bounds, at the price of indexes several
times larger than the data. Approximate indexes, such as
\seismic~\cite{bruch2024seismic,nardini2027compact} and
graph indexes like HNSW~\cite{malkov2020hnsw,delfino2025kannolo}, give up
the guarantee of exact results, and trade recall for speed through
parameters tuned for every collection, model and depth $k$. The two
families have been compared at length, and the received view is that
approximate indexes are faster than exact
ones~\cite{bruch2024seismic,nardini2027compact,delfino2025kannolo,simhadri2025bigann23}.

This paper challenges that view with classical IR alone. We present \abc,
an exact index for LSR with no parameters, built only from standard
techniques: term-at-a-time (TAAT) evaluation~\cite{turtle1995query},
Block-Max MaxScore~\cite{chakrabarti2011interval,dimopoulos2013bmm}, binary
packing~\cite{lemire2015decoding} and doc-id
reordering by graph
bisection~\cite{dhulipala2016compressing}. What makes
the difference is how they are combined and engineered for modern CPUs.
Take TAAT, considered in the research literature inferior to
document-at-a-time (DAAT) evaluation because its accumulators do not fit in
cache~\cite{fontoura2011evaluation}. Cut the doc-id space into aligned
blocks whose accumulators stay in the
L1 cache, and store each posting as a 16-bit offset that is directly its
accumulator slot: scoring a posting then needs no decoding and no
comparison, only a multiply-add into a slot that is already in cache.

On top of this loop, dynamic pruning over posting lists pays off again,
wacky weights notwithstanding. Our \abcbmms runs Block-Max MaxScore inside
every block, with maxima aligned across terms, so
that the bounds of the query terms add up, and with a cost model that
decides, block by block, which lists to read in full and how to complete
the few candidates that may still enter the top $k$. Pruning makes it
always faster than the exhaustive \abc, by up to 4.1$\times$, whereas
\lucene's Block-Max MaxScore scorer, which also prunes window by window, is
almost always slower than \lucene's own exhaustive evaluation in our
experiments.

Our indexes compute every score
from the stored weights: with 8-bit weights, as in the other block-based
indexes, they lose less than 1\% of recall, and with the original 32-bit
weights they return the exact top-$k$, for at most 40\% more time than
with 8-bit weights. An
\texttt{f32} \abc index, built on the fly, thus also computes the exact
ground truth of a query set, the reference against which LSR models and
approximate indexes are measured, 27--52$\times$ faster than a
brute-force scan. There is nothing to tune: every model is
indexed the same way, in seconds, and one index serves every depth $k$,
whereas \seismic needs a grid search and a different index for every $k$,
and HNSW takes hours to build.

We evaluate our indexes on MS MARCO encoded with four LSR models, from the
historical CoCondenser to the recent OS-gte, SparseUp and SPLARE, against
the best exact and approximate indexes. Our contributions are the
following.
\begin{itemize}
  \item \emph{Exhaustive TAAT over cache-resident blocks, built in seconds
    (\abc).} Our index turns TAAT, believed to lose to DAAT on cache
    misses, into a loop that stays in the L1 cache, with a compressed index smaller than those of \qdrant and \lucene. The
    doc-id reordering, which makes the index smaller, no longer dominates the build: our graph
    bisection reorders the collection 19.5$\times$ faster than
    \texttt{rgbisect}~\cite{mackenzie2021faster}, with compressed doc-ids
    only 1.2\% larger.
  \item \emph{Block-Max MaxScore inside the TAAT loop: the fastest exact
    index (\abcbmms).} \lucene also prunes posting lists window by window
    with Block-Max MaxScore, and there pruning slows it down. Inside our
    blocks it pays off: \abcbmms is 13--55$\times$ faster than \lucene and
    6.5--84$\times$ faster than the sparse index of
    \qdrant~\cite{qdrant2025sparse}, and the fastest exact index on all
    four datasets and at every $k$; against \bmp it is 2.9--18$\times$ faster with an
    index 5--7$\times$ smaller.
  \item \emph{Exact, yet competitive with approximate indexes.} At $k=10$,
    \seismic and HNSW at 99\% recall are up to 4.3$\times$ faster than
    \abcbmms, with indexes 1.5--3.6$\times$ larger, but their cost grows faster
    with $k$. At $k=100$ they are at most 2.4$\times$ faster, and \abcbmms is
    the fastest on OS-gte, where HNSW does not reach 99\% recall, and faster
    than \seismic on SparseUp. At $k=1000$ \abcbmms is the fastest index on
    all four datasets, 1.7--7.4$\times$ faster than \seismic, and it is built
    in 13--32\,s instead of the 9--27 hours of HNSW. Exactness and space succinctness come with the extra
    benefit that no parameter tuning is needed.
\end{itemize}

\section{Background and Related Work}
\label{sec:related}

\paragraph{Learned sparse retrieval.} LSR models encode queries and
documents as sparse vectors over the vocabulary of a language
model~\cite{formal2021splade,formal2022distillation}.
Relevance is the inner product, so retrieval is top-$k$ maximum
inner-product search over sparse vectors, and the natural data structure is
the inverted index, whose posting lists can be compressed with the
techniques surveyed by Pibiri and Venturini~\cite{pibiri2020techniques}. Learned weights break the assumptions of classic query
processing: queries have tens of terms, weights are large on frequent
terms, and documents are expanded with terms they do not contain. Mackenzie
et al.\ call them ``wacky weights'', on which ``few documents can actually
be skipped if exact query evaluation is desired''~\cite{mackenzie2021wacky}.

\paragraph{DAAT and dynamic pruning.} Document-at-a-time (DAAT) processing
advances one cursor per query term in doc-id order and scores one document at
a time. Dynamic pruning skips the documents that cannot enter the top $k$,
using per-term upper bounds. MaxScore~\cite{turtle1995query} splits the
terms into \emph{essential} ones, read in full, and \emph{non-essential}
ones, whose total contribution is too small to bring a document into the
top $k$ on its own. The non-essential terms are read only for the few
documents that may still enter it. WAND~\cite{broder2003wand} pivots on the
prefix sums of the bounds, and Block-Max WAND~\cite{ding2011bmw} adds
per-block maxima. Block-Max
MaxScore~\cite{chakrabarti2011interval,dimopoulos2013bmm} runs MaxScore inside doc-id intervals, with tighter bounds for each
interval. Dimopoulos et al.~\cite{dimopoulos2013bmm} already compute
these bounds over fixed ranges of doc-ids whose size is a power of two, as
\abcbmms does (Section~\ref{sec:bmms}). Anytime ranking visits ranges of
doc-ids in decreasing order of their bound~\cite{mackenzie2022anytime}.

\paragraph{Block-based pruning for LSR} \bmp~\cite{mallia2024bmp}
partitions the reordered doc-id space into blocks of a few documents, stores
8-bit block maxima per term and, per block, inverted lists of the block's
postings, and scores the blocks by decreasing upper bound until the
threshold exceeds the next bound. Each block is scored term at a time, into one accumulator per
document of the block.
\spidx~\cite{carlson2025sp} groups blocks into superblocks pruned before
their blocks, and \lsp~\cite{carlson2026lsp} visits the most promising
superblocks with 4-bit maxima and a forward index. In their rank-safe
settings they are an order of magnitude faster than the MaxScore of \pisa~\cite{mallia2019pisa} on
SPLADE, but their indexes are several times larger than the data.
\rise~\cite{savino2026rise}, a Rust re-implementation of
DS2I~\cite{ottaviano2014partitioned} about as fast as \pisa, offers the
same document-at-a-time algorithms, MaxScore and Block-Max MaxScore among
them.

\paragraph{Approximate retrieval.} \seismic~\cite{bruch2024seismic,nardini2027compact}
keeps the top entries of each posting list, clusters them into blocks
summarized by quantized max-vectors, skips the blocks whose summary score is
below a fraction of the threshold, and scores the others with a forward
index. Graph indexes such as HNSW~\cite{malkov2020hnsw}, implemented for
sparse vectors in \kannolo~\cite{delfino2025kannolo} and in the winners of
the BigANN'23 sparse track~\cite{simhadri2025bigann23}, are the other family.
Both trade recall for speed through parameters chosen by a grid search on
each collection~\cite{bruch2025scalability}.

\paragraph{Term-at-a-time.} Term-at-a-time (TAAT) processing adds
one posting list at a time into an accumulator per document. DAAT came to dominate research~\cite{tonellotto2018efficient}. On large
memory-resident indexes, the cache misses of TAAT increase drastically,
which Fontoura et al.\ attribute to ``random access over the large array of
accumulators''~\cite{fontoura2011evaluation}. The research literature studies
TAAT with accumulators over the whole collection or over a pruned subset of
it~\cite{fontoura2011evaluation}. Cache-sized windows of doc-ids appear in
DAAT, but only to hold a bitmap of live blocks~\cite{dimopoulos2013filter}.
Search engines took another road: \lucene scores disjunctions term at a time in small windows of
doc-ids, whose accumulators stay in
cache~\cite{lucene_booleanscorer}, and so does
\qdrant~\cite{qdrant2025sparse}. The compressed sparse index of \qdrant scores every posting, in windows that start at the smallest
current doc-id of the query lists, so every posting must be rebased to its
window. In \abc, instead, the blocks are fixed and aligned, and the 16-bit
offset of a posting is directly its accumulator slot.
\lucene's Block-Max MaxScore scorer prunes the windows, splitting the
clauses anew in every window and completing the candidates term at a
time~\cite{lucene_maxscorebulkscorer}.
Our \abcbmms also prunes window by window, but differs in three choices. The offset of a posting is its accumulator
slot. The 8-bit maxima, one for every 1024 doc-ids, are aligned across
terms, so that the bounds of the query terms add up (\lucene's cover chunks
of postings, so each list's bound must be recomputed for every window). A
cost model decides which lists to skip, and whether to probe the few
candidates in a list or to add the list whole
(Section~\ref{sec:bmms}). To our knowledge, no published study measures
TAAT over cache-sized windows of global posting lists, with or without
pruning, against DAAT, block-based or approximate LSR
indexes.

\section{The Exhaustive Index \abc}
\label{sec:abc}
Classic TAAT keeps one accumulator per document: each posting list is read
sequentially, but its updates land at random positions of an array as large
as the collection, and on large indexes the cache misses grow sharply~\cite{fontoura2011evaluation}. \lucene and \qdrant avoid these misses
by evaluating the query on one small window of doc-ids at a time, whose
accumulators stay in cache. Both use what we call \emph{soft windowing}: the
window exists only at query time, and for each term the search reads
postings until the next one falls outside the window. The search must thus find where the window ends in the list of each term and bring every doc-id to its position within it.
Soft windowing is usually paired with a partitioning of the posting lists by
cardinality, for example into chunks of a fixed number of postings, none of
which corresponds to a window.

Our index \abc uses instead what we call \emph{hard windowing}: the windows are fixed, aligned intervals of the doc-id space, the same for every term, and they
are also the blocks in which every posting list is stored. The postings that a term has in a window are thus known in advance, and the search
does not have to find where the window ends. A posting does not
store its block either, only its position within it, which is already its
slot in the accumulator. In exhaustive search, this keeps the work per
posting simple for the CPU: no comparison with the window and no shift, only
its contribution added to its slot. This already speeds up exhaustive search,
and hard windows will be crucial for dynamic pruning (Section~\ref{sec:bmms}).

Hard windows cost negligible space: stored with known techniques, the compressed index of \abc is smaller than that of \qdrant (Section~\ref{sec:u16}), and its
search is several times faster than that of \qdrant with the same weights and
an order of magnitude faster than that of \lucene (Section~\ref{sec:qdrant}). All the measurements of this section are on MS
MARCO~\cite{nguyen2016msmarco} encoded with the CoCondenser-based SPLADE
model~\cite{gao2022cocondenser,formal2022distillation}
(Section~\ref{sec:exp-settings}).

\subsection{Term-at-a-time over hard windows}
\label{sec:taat}
In TAAT over windows, the window is the range of doc-ids covered by the
array of accumulators, and it is convenient to make it coincide with the
blocks into which the doc-id universe is partitioned.
The universe is cut into blocks of $B$ consecutive doc-ids, the same for all
the terms, and the array has one slot per doc-id of a block. In this way, a
posting stores only its offset within its block, since the list is stored
block by block; its doc-id is $\mathit{block} \cdot B + \mathit{offset}$,
and its offset is directly the index of its slot.

The width $B$ thus fixes the size of the accumulator, and we choose it for
the accumulator to stay in the L1 cache. We fix $B = 2^{13}$, the fastest
value in our experiments: an accumulator of 32-bit floats takes 32\,KiB and
fits in the 48\,KiB L1 data cache of our machine, and an offset fits in
16 bits.

A query is evaluated one block at a time, in ascending order, visiting only
the blocks in which its terms have postings. Within a block, the query terms
are taken one after the other, and the contribution $q_t d_t$ of each of
their postings is added to its slot, where $q_t$ and $d_t$ are the weights of
term $t$ in the query and in the document. Once all the terms are done, the
accumulator holds the exact score of every document of the block. A last pass
over the array of accumulators then moves these scores to a heap that keeps the best $k$ documents seen so
far, and prepares the accumulator for the next block. The threshold of the
heap is $-\infty$ until the heap is full, and then the lowest score it holds.

This last pass reads every slot of every visited block, so it must cost
little. In its simplest form, it offers each slot to the heap, which accepts
it only if it exceeds the threshold, and it resets the slot to zero as it
reads it, so the accumulator needs no separate clearing. No slot is tested
against zero: documents with a zero score are discarded when the results are
collected.

Every slot still costs a comparison with the threshold, and SIMD instructions
make it cheaper. An AVX2 register holds 8 scores of 32 bits, so four SIMD
comparisons check 32 slots at once, with a single branch. Only when this
branch is taken does the scan look for the groups of 8 that hold a score
above the threshold, and hand their slots to the heap. Once the heap is full
the branch is almost never taken, so it is well predicted. Each slot is still
reset in the same pass.

\subsection{The \abc memory layout}
\label{sec:u16}
\begin{figure*}
\centering
\input{fig_u16}
\Description{A term's posting list over six blocks of 8192 doc-ids, its
used-block bitvector 101101, the array of per-block starts pointing into the
arrays of 16-bit offsets and 8-bit weights, and the dense block 3 stored as
8192 weights, one per doc-id.}
\caption{The memory layout of \abc on one term $t$, whose postings fall in
blocks 0, 2, 3 and 5: blocks 0, 2 and 5 are in sparse format, and block 3 is in dense format.}
\label{fig:u16}
\end{figure*}
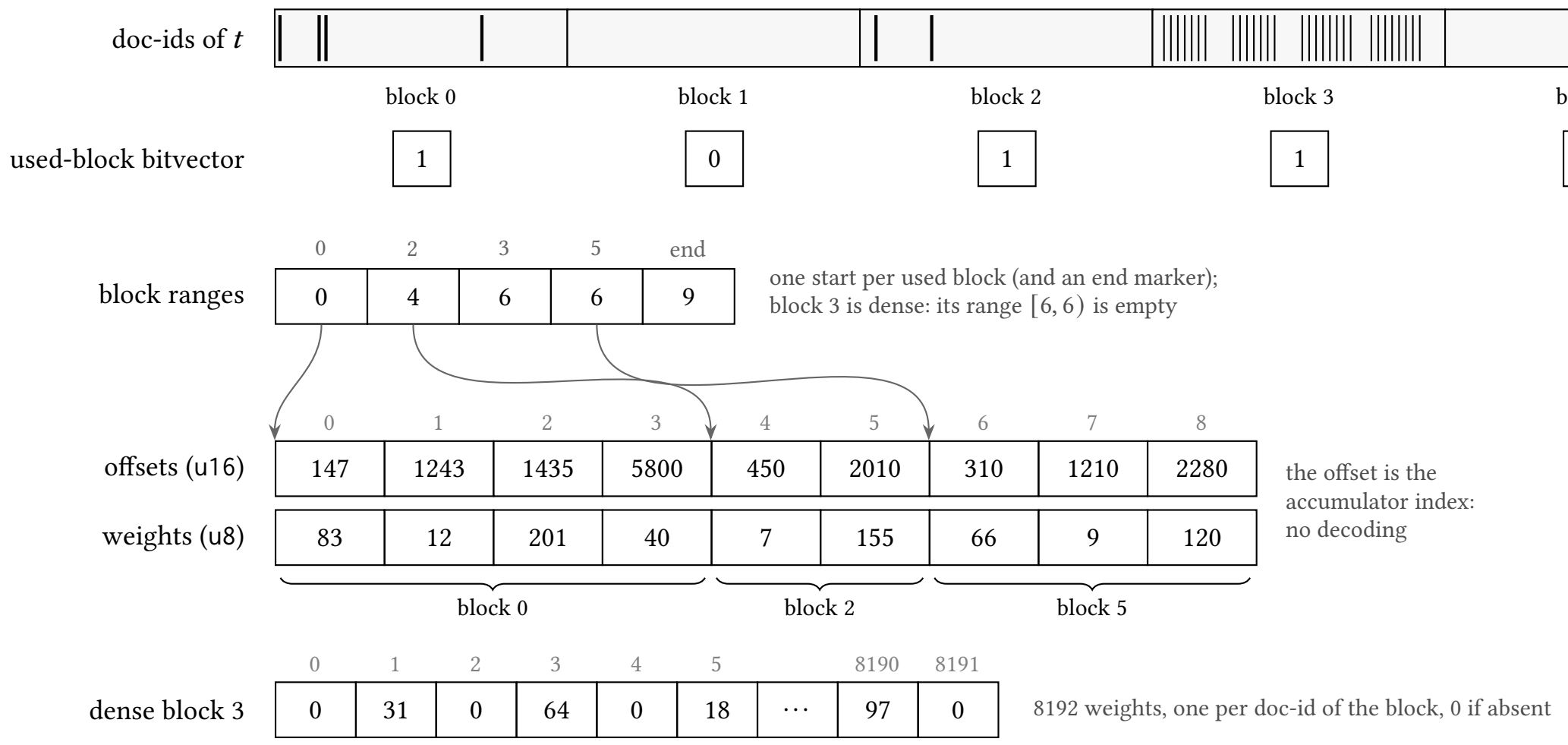

The representation of a posting list consists of three parts: the blocks in
which the term has postings, the offsets of its postings within each of
these blocks, and their weights. It is inspired by the elegant
representation of Elias and Fano~\cite{elias1974efficient,fano1971number}, from which
quasi-succinct indices~\cite{vigna2013quasi}, partitioned
Elias-Fano~\cite{ottaviano2014partitioned} and Roaring
bitmaps~\cite{chambi2016roaring} also drew inspiration: every value is
divided into a high part, which selects a bucket of the universe, and a low
part, its position within the bucket. In \abc the high part is the block,
implicit in the structure of the list, and the low part is the offset.
Figure~\ref{fig:u16} shows the layout on one term. We first describe the
layout that stores the offsets uncompressed, which is the fastest, and then
its compressed variant.

\paragraph{Used blocks.} To know which blocks to visit, each term has a
bitvector with one bit per block, set when the term has postings in the
block, and the bitwise OR of the bitvectors of the query terms gives the
blocks to visit. The bitvectors take negligible space: with CoCondenser, about 4\,MiB in all, 0.1\% of
the index of \abc.

\paragraph{The postings of a block.} Each block of a posting list is stored
as the offsets of its postings, sorted, in 16 bits each, and their weights
in the same order. During a query, the blocks are
visited in ascending order and each posting list is read from left to
right, so every offset that is read belongs to the block being visited. The
bitvector of the term tells whether its list has postings in that block, and
a small array of \emph{block ranges} stores where every block of a list
begins, so its length is the distance between its start and the next one.
Each term thus reads its list with a \emph{cursor} that only moves forward,
and reaching the next block requires no search. Adding a block of a list to
the accumulator is a tight loop over its offsets and weights, with no
branches on the data. In Figure~\ref{fig:u16}, the cursor of $t$ reads the
four postings of block 0, adding the contribution of weight 83 to slot 147,
skips block 1, and reads the two postings of block 2 and the three of
block 5; block 3 is a different case, which we discuss below.

\paragraph{Weights.}
\label{sec:weights}
The contribution of a posting comes from its weight, stored as an 8-bit
integer (\texttt{u8}), a 16-bit float (\texttt{f16}) or a 32-bit float
(\texttt{f32}). With 8 bits, each posting list has its own affine
quantization $d_t \approx m_t + r\,s_t$, where $r$ is the stored integer.
The query weight is folded into the two parameters of the list once per
query, $q_t d_t \approx (q_t m_t) + r\,(q_t s_t)$, so each posting costs a
single multiply-add. With 32-bit weights the stored weights are the original ones, so the search
returns the true top-$k$, and with 8-bit
weights it loses less than 1\% of recall on all our models. On CoCondenser,
16-bit weights lose about 0.1\%, and against 8-bit weights \abc takes
2--5\% more time with 16-bit weights and 36--40\% more with 32-bit ones
(Table~\ref{tab:qdrant}).

\paragraph{Dense blocks.} The offsets and weights described so far are the
\emph{sparse format} of a block of a list. Storing the offsets becomes costly
when the block is dense. The \emph{density} of a block of a list is the fraction
of the doc-ids of the block in which the term occurs. A dense block is
therefore stored in \emph{dense format}: one weight for every doc-id of the
block, 0 for the documents without the term, and no offsets. The weight zero is free to
mark the absent documents, because 8-bit weights are quantized to the levels
$1,\dots,255$ and a 16- or 32-bit weight equal to zero contributes nothing;
a dense block is thus added to the accumulator with one sequential,
vectorized pass. In Figure~\ref{fig:u16}, block 3 is stored in dense format. A block is stored in dense format when its density
makes this format no larger than offsets and weights: 1/2 with 16-bit
weights and 2/3 with 32-bit ones. With 8-bit weights the threshold is 1/4 instead of 1/3; between the two,
time and space change by less than 1\%. Dense blocks become more frequent when the doc-ids are reordered so that documents that often occur in the same posting lists are close. We use the recursive graph bisection of
Dhulipala et al.~\cite{dhulipala2016compressing}, in the faster version of
Mackenzie et al.~\cite{mackenzie2021faster}, which we improve in
Section~\ref{sec:bp}. The postings of a term
then gather in fewer, denser blocks, and this is where the dense format pays
off. On CoCondenser with 8-bit weights, 0.4\% of the blocks of lists are
dense, and they hold 23\% of the postings, against 0.02\% without
reordering. The dense format saves 4.8\% of the space of \abc
and 10--11\% of its query time, while without reordering it changes
neither. Quasi-succinct indices, partitioned Elias-Fano and Roaring bitmaps also
switch to a dense representation where the values are dense; our dense format, in
addition, carries the weights.

\paragraph{Compressed layout.} In the sparse format, every offset still takes
16 bits. \emph{Binary packing}~\cite{lemire2015decoding} compresses a sorted
sequence by cutting the gaps between consecutive values into groups, usually
of 128, and storing each group with the number of bits of its largest gap.
In the compressed variant of the index, \abcpacked, each block of a list is
stored as a header with its first offset and its length, 16 bits each,
followed by its gaps, bit-packed in groups of 8 instead of 128, because
within a block a term has few postings. A
group is decoded with a handful of SIMD instructions and no branches, and
the search reads the blocks of a list in the same order, with the same
forward-only cursor. In \abcpacked, dense blocks are stored in sparse format too, because their gaps are small and take a few bits. With the threshold of \abc, the dense
format would make the compressed index 13\% larger.

\paragraph{The cost of hard windows.} Besides the bitvectors, hard windows
add to the index only where each block of a list ends. In \abc this is a
4-byte entry in the block ranges. In \abcpacked it is the 16-bit length in the header of the block; the other half of the header, the first offset, belongs to a posting. On CoCondenser, a block of a list holds 44 postings on average, and 65
with reordered doc-ids, so this information takes less than 0.1 bytes per
posting and 2--3\% of the index, depending on the layout and on reordering:
negligible space. Even with it, the index of \abcpacked is smaller than
that of \qdrant and, with reordered doc-ids, than that of \lucene, whose
index is reordered too.

\subsection{Comparison with \qdrant and \lucene}
\label{sec:qdrant}
We compare \abc with the two exhaustive engines that use soft windowing: the
sparse index of \qdrant~\cite{qdrant2025sparse} and the exhaustive
evaluation of \lucene~\cite{lucene_booleanscorer}. \qdrant v1.20.2 is a
production engine, and its sparse index is a Rust library that we call from
our own benchmark, with the same compiler optimization settings and the same
weight precisions as \abc: 8, 16 and 32 bits. \lucene 10.5.2 evaluates disjunctions term at a time
in windows of 4096 doc-ids, with 8-bit weights and its own quantization
strategy, on an index whose
doc-ids are reordered by its own graph bisection, BP
(Section~\ref{sec:exp-settings}).
Table~\ref{tab:qdrant} reports query time, space, build time and recall on
CoCondenser.

\begin{table*}
\caption{Exhaustive search on CoCondenser: latency in \textmu s/query on one
thread, index size in MiB, build time on 20 threads, and recall@$k$ against
the exact \texttt{f32} top-$k$. The best value of each weight type in bold. The rows of \abc are not reordered; the build
of \lucene includes its BP.}
\label{tab:qdrant}
\begin{tabular}{llrrrrrl}
\toprule
Weights & Index & $k=10$ & $k=100$ & $k=1000$ & Size & Build & Recall@10/100/1000 \\
\midrule
\multirow{5}{*}{u8} & \qdrant v1.20.2 & 30,531 & 30,744 & 30,692 & 2,289 & 35.5\,s & .9925/.9953/.9963 \\
 & \abc & \textbf{4,865} & \textbf{4,894} & \textbf{5,068} & 3,131 & 6.3\,s & .9925/.9953/.9964 \\
 & \abcpacked & 6,385 & 6,219 & 7,498 & 2,141 & 5.2\,s & .9925/.9953/.9964 \\
 & \abcpacked, reordered & 5,850 & 6,050 & 6,233 & \textbf{1,814} & 11.9\,s & .9925/.9953/.9964 \\
 & \lucene 10.5.2, reordered & 59,386 & 59,022 & 60,934 & 1,941 & 523\,s & .9898/.9940/.9953 \\
\midrule
\multirow{4}{*}{f16} & \qdrant v1.20.2 & 29,800 & 29,919 & 29,972 & 3,299 & 40.9\,s & .9990/.9994/.9996 \\
 & \abc & \textbf{5,083} & \textbf{5,006} & \textbf{5,333} & 4,142 & 6.7\,s & .9989/.9994/.9996 \\
 & \abcpacked & 6,404 & 6,553 & 6,600 & 3,152 & 5.4\,s & .9989/.9994/.9996 \\
 & \abcpacked, reordered & 6,642 & 6,425 & 6,447 & \textbf{2,825} & 11.9\,s & .9989/.9994/.9996 \\
\midrule
\multirow{4}{*}{f32} & \qdrant v1.20.2 & 29,990 & 30,047 & 30,652 & 5,319 & 57.1\,s & 1/1/1 \\
 & \abc & \textbf{6,833} & \textbf{6,681} & \textbf{6,873} & 6,165 & 8.9\,s & 1/1/1 \\
 & \abcpacked & 8,094 & 8,101 & 8,959 & 5,175 & 5.8\,s & 1/1/1 \\
 & \abcpacked, reordered & 8,180 & 8,364 & 8,366 & \textbf{4,849} & 12.3\,s & 1/1/1 \\
\bottomrule
\end{tabular}
\end{table*}

With the same weight precision, \abc is 4.4--6.3$\times$ faster than
\qdrant, and it is about 12$\times$ faster than \lucene. Its compressed variant \abcpacked is 3.4--5.2$\times$ faster
than \qdrant, with a smaller index, with or without reordering. With reordered doc-ids, \abcpacked is
also about 10$\times$ faster than \lucene, with an index 6.5\% smaller. The index of \abc is 16--37\% larger than that of
\qdrant.

\paragraph{Space.} \qdrant and \lucene divide each posting list into chunks
of a fixed number of postings, 128 in \qdrant and 256 in \lucene, and
bit-pack all the gaps of a full chunk with a single width, whereas
\abcpacked uses one width for every 8 gaps. Against
\qdrant, whose 8-bit weights also take one byte per posting, the difference
lies in the doc-ids, which \abcpacked packs with finer widths. Everything but the weights takes 10.1 bits per posting
in \qdrant, 8.9 in \abcpacked and 6.3 with reordered doc-ids, while \abc
stores each offset in 16 bits. Against \lucene, which also compresses its
weights, only the total sizes can be compared.
With 8-bit weights, reordering makes the index of \abcpacked 15\% smaller.
It helps \qdrant too: on the reordered collection its index is 10\% smaller
and its search 15\% faster. \lastedit{}With both reordered, \abcpacked is about 4$\times$
faster than \qdrant, with an index 12\% smaller.

\paragraph{The work of soft windows.} At query time, soft windows cost \qdrant and \lucene work that hard windows remove. For every window and every term, they must
find where the window ends in a chunk and bring every doc-id to its slot. The soft windows of \qdrant are also \emph{floating}: a window starts
at the smallest doc-id not yet read among the lists of the query. \qdrant
finds where the window ends in a chunk with a binary search, and subtracts
the start of the window from every doc-id. \lucene finds the end of the
window by searching the chunk for the first doc-id past it, and brings every
doc-id to its position within the window. Neither is needed in \abc, where
a window is a block of the index. On the uncompressed \abc without the dense
format, replacing the hard windows with floating windows like those of
\qdrant, with the same sum of the contributions, the same scan and the same
heap, makes the search about 30\% slower.

\paragraph{The inner loop.} By inner loop we mean the work that every window
repeats: adding the contribution of each posting to its slot, and scanning
the accumulator at the end. Two choices in this loop also count, and they do
not depend on the windows: how the contribution of a posting is
computed, and how the accumulator is scanned. In \abc each posting costs one
multiply-add straight into its slot. \qdrant and \lucene instead go through
the postings of a chunk twice, with a buffer in between. \qdrant first
dequantizes the weights of the chunk into the buffer, and then multiplies
each of them by the query weight and adds it to its slot. \lucene first
computes the score of every posting of the chunk into the buffer, applying
its similarity to the stored weights, and then adds each score to its slot.
As for the scan, \abc in its simple form offers each slot
to the heap and resets it in the same pass (Section~\ref{sec:taat}).
\qdrant instead zeroes its array of accumulators before scoring a window,
and then tests every slot with a branch that depends on the data, keeping
the score if it is non-zero and above the threshold. We ported our simple scan into \qdrant, and it made \qdrant about 2.5 times
faster. Folding the query weight into the quantization of \qdrant, instead,
did not make it faster. However, our version of \qdrant is still
1.4--1.5$\times$ slower than \abcpacked with the same scan. This is probably
because of soft windowing, which alone makes \abc about 30\% slower.

\section{Dynamic Pruning with Block-Max MaxScore}
\label{sec:bmms}

Exhaustive \abc scores every document of every block that contains a query
term, although only the best $k$ matter. Dynamic pruning would like to skip this work, using upper bounds on the
scores, but on learned sparse representations it is
known to pay off little, or even to hurt. On SPLADE, WAND and Block-Max WAND
are slower than scoring every
posting~\cite{mackenzie2021wacky,mackenzie2023efficient}, and MaxScore is
slower than an exhaustive score-at-a-time evaluation~\cite{mallia2024bmp}.
The two production-ready engines of Section~\ref{sec:qdrant} confirm it. \lucene prunes with
Block-Max MaxScore, window by window~\cite{lucene_maxscorebulkscorer}, and
in our experiments it is almost always slower than its own exhaustive
evaluation, by up to 2.7$\times$ (Section~\ref{sec:exp-compare}). \qdrant
prunes with WAND only in its uncompressed in-memory index, which with
32-bit weights is 1.1--1.4$\times$ slower than its compressed index without
pruning on all four datasets.

This section turns \abc into \abcbmms, which prunes with Block-Max MaxScore,
and there pruning makes the search faster. On our four datasets \abcbmms is always faster than the exhaustive \abc, both with reordered
doc-ids, from 1.1$\times$ at $k=1000$ up to 4.1$\times$ at $k=10$ (Table~\ref{tab:abc}). Pruning inside the term-at-a-time loop relies on the hard windows. Since they
are the same for every term, an upper bound
on the contributions of a term in an interval of a block refers to the same
doc-ids for all the terms, so the bounds of the query terms add up, as the block maxima of \bmp
do~\cite{mallia2024bmp}. The maxima of \lucene cover chunks of postings,
which differ from list to list, so in every window \lucene must first derive
the bound of every list from the chunks that overlap it.

\subsection{One byte per sub-block}
\label{sec:bmms-layout}
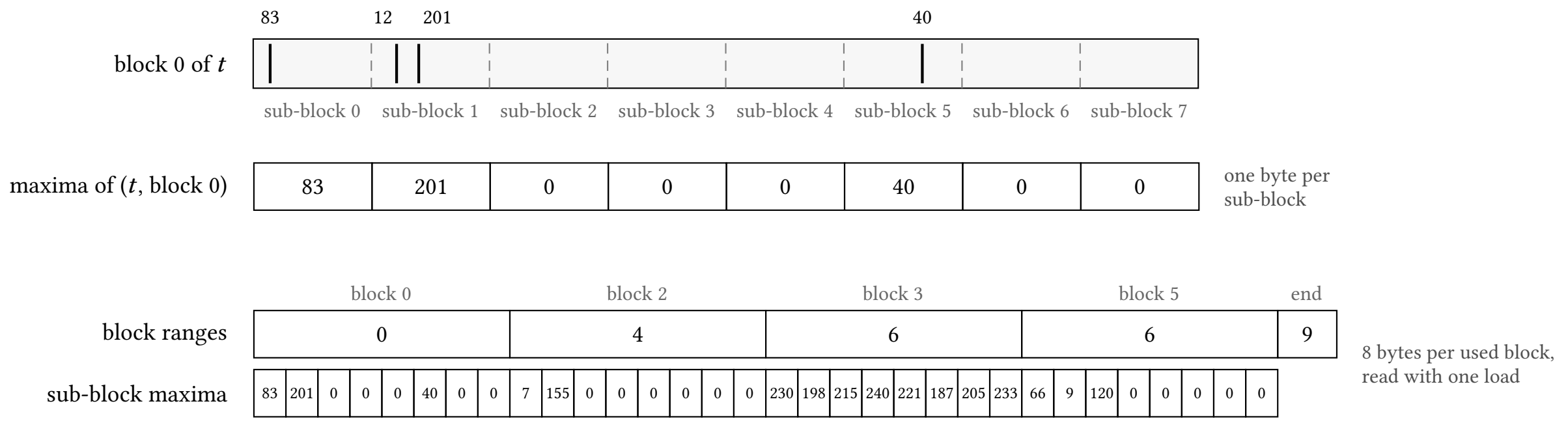
\begin{figure*}
\centering
\input{fig_bmms}
\Description{The sub-block maxima of a term: block 0 split into 8 sub-blocks of 1024 doc-ids, each with the largest weight of the term in it.}
\caption{The sub-block maxima of \abcbmms on the term $t$ of
Figure~\ref{fig:u16}: block 0 is split into 8 sub-blocks of 1024 doc-ids,
each with the largest weight of $t$ in it, 0 when $t$ does not occur
there.}
\label{fig:bmms}
\end{figure*}

To bound the scores, each block of 8192 doc-ids is divided into 8
\emph{sub-blocks} of 1024 doc-ids, fixed and aligned like the blocks. For
every block of a list, the index stores the largest weight of the term in
each of its 8 sub-blocks, quantized to one byte and rounded up, and 0 for a
sub-block without postings of the term. The 8 maxima are stored next to the
block ranges, so the cursor that reaches the postings of a block of a list
also reaches its maxima (Figure~\ref{fig:bmms}). Nothing else changes:
postings, used-block bitvectors and dense blocks are those of \abc.

The maxima are stored only for the blocks of a list that its bitvector
marks, 8 bytes each, whatever the number of their postings. On CoCondenser they take 124\,MiB, 4\% of the index. Doc-id reordering keeps this cost low, because it gathers the
postings of a term in 32\% fewer blocks; without it, the maxima would take about
184\,MiB.

\bmp~\cite{mallia2024bmp} keeps one 8-bit maximum for every term and every
block of $b$ documents, zero meaning absent. Blocks of 8 or 16 documents give
tight bounds, and on CoCondenser their maxima take 30.0 and 15.0\,GB at
$b=8$ and 16, and 5.5 and 4.1\,GB
compressed~\cite{mallia2024bmp}; \lsp reports 7.6\,GB for a sparse layout at
$b=8$~\cite{carlson2026lsp}. Our maxima cover 1024 documents instead of 8 or
16, and the finer pruning happens inside the blocks
(Section~\ref{sec:bmms-maxscore}).

\subsection{Visit the promising blocks first}
\label{sec:bmms-order}
Before reading any posting, the search sums the maxima of the query terms.
One ascending pass over the blocks of the query lists, the same pass that
exhaustive \abc makes, adds to each sub-block the query weight of every term
times its maximum there. The result is the \emph{bound} of the sub-block,
and the bound of a block is that of its best sub-block. The same pass records where the postings of every list begin in every
block, so the blocks can be visited in any order. The blocks are then visited in decreasing order of their bound, as in \bmp~\cite{mallia2024bmp}; with
1,080 blocks on MS MARCO, sorting them takes negligible time. The first blocks fill
the heap with documents of high score, so the threshold $\theta$, the $k$-th
best score found so far, rises quickly. As soon as the bound of the next
block falls below $\theta$, no document in it or in any later block can
enter the top $k$, and the search stops. Inside a visited block, the
sub-blocks whose bound is below $\theta$ are skipped.

The main effect of this order is a high $\theta$ from the first blocks on.
Most blocks are still visited, 626 of the 1,080 on CoCondenser at $k=10$,
because queries have 44 terms on average and most blocks hold a sub-block whose bound exceeds $\theta$. Most of the work must therefore be saved inside the blocks.

\subsection{MaxScore, one block at a time}
\label{sec:bmms-maxscore}
Inside a visited block, the search applies MaxScore to the lists of the query. The \emph{bound of a
list} in the block is the query weight of its term times its largest maximum
over the sub-blocks not skipped. The lists whose bounds
sum to less than $\theta$ can be declared \emph{non-essential} in the
block, because a document that occurs only in them cannot enter the top $k$.
The \emph{essential} lists are added whole to the accumulator, as in
exhaustive \abc. A single vectorized pass over the accumulator then keeps as
\emph{candidates} only the documents whose partial score, plus what the
non-essential lists can still add in their sub-block, can reach $\theta$.
The test uses the maxima of the lists in the sub-block of the document,
which are tighter than the bounds of the lists. All the other documents are dropped without reading a single non-essential
posting. Finally, as in the term-at-a-time MaxScore of Turtle and
Flood~\cite{turtle1995query,fontoura2011evaluation} and in \lucene, the
non-essential lists are added to the candidates one at a time, and after each list a candidate is dropped when the lists still
missing can no longer bring it to $\theta$. The candidates that remain after the last list enter the heap with their
exact scores, and when none remains earlier, the other lists are not read
at all.

The split is redone in every block, with the current $\theta$ and the
bounds of the lists in that block, so a term with large weights in a few
blocks is essential there and non-essential everywhere else. Term-at-a-time
processing makes this cheap, because the split is decided once per block of
8192 documents and the candidates are slots of an accumulator of 32\,KiB
that stays in cache. \lucene also redoes the split in every
window~\cite{lucene_maxscorebulkscorer}, but it first derives the bound of every list from the maxima of the chunks
that overlap the window, while in \abcbmms the bound of a list is the largest of its maxima in the
sub-blocks not skipped. On CoCondenser at $k=10$, the
split inside the blocks makes the search 3.0$\times$ faster than block and
sub-block pruning alone.

\paragraph{Exactness.} A block, a sub-block or a document is dropped only
when an upper bound of its score is below $\theta$, and $\theta$ never
decreases, so the result is the exact top-$k$ of exhaustive \abc, up to ties.
The maxima are rounded up, so a bound is never below the weights it covers,
and $\theta$ is lowered by a relative slack of $10^{-4}$ that absorbs the
different summation orders of bounds and scores. Query weights are assumed
non-negative, as in SPLADE; with a negative weight the bounds would not hold.

\subsection{Choosing what not to read}
\label{sec:bmms-select}
Any set of lists whose bounds sum below $\theta$ can be non-essential, so
the split is a choice, and it decides how much a block costs. Two things
can be chosen: the threshold below which the bounds of the non-essential
lists must sum, and the lists themselves.

Any threshold $\tau \le \theta$ keeps the search exact, and all the
implementations of MaxScore we know of use $\tau=\theta$. A lower $\tau$ seems
wasteful, because fewer lists can be non-essential, but it can pay off.
Take a block with $\theta=10$ and three lists: A, essential, with bound 6,
and two long lists B and C, with bounds 5 and 4.9. With $\tau=\theta$, B
and C are both non-essential, and a document needs a partial score of only
0.1 from A to be a candidate. Thousands of documents are then candidates,
and B and C end up read almost whole. With $\tau=5$, B is essential and
read whole, but a candidate needs a partial score of at least 5.1 from A
and B, so the candidates are few and C is read only for them. In \abcbmms,
$\tau$ depends on how far the block lies above $\theta$. When the bound of
the block barely exceeds $\theta$, few of its documents can reach it, and
$\tau$ can be $\theta$ itself. When the block lies far above $\theta$,
many of its documents may score high, and a smaller $\tau$ keeps the
candidates few. We set $\tau$ from the ratio between $\theta$ and the bound
$\beta$ of the block, as
$\tau = \theta \cdot \min(1, \max(0.4,\, \theta/\beta + 0.25))$. On
CoCondenser, our $\tau$ cuts the candidates to a third at $k=10$ and
makes the search 9--15\% faster than $\tau=\theta$, more at larger $k$.

Below $\tau$, choosing the lists is a knapsack problem. What a
non-essential list saves is its postings, which are not added to the
accumulator, while its bound adds to the sum that must stay below $\tau$.
A list with a small bound and few postings saves little, and a long list of
small weights saves a lot, so the goal is to leave out as many postings as
possible, keeping their bounds below $\tau$. Textbook MaxScore, \rise and
\pisa take as non-essential the terms with the smallest bounds. \lucene
sorts its clauses by a ratio of bound to cost, the bound of a clause in the
window over the length of its whole list~\cite{lucene_maxscorebulkscorer},
and takes them greedily. We solve the knapsack with the same kind of greedy
choice, but our ratio counts only the postings of the list in the block:
the lists are sorted by their bound per posting in the block, and the
non-essential ones are chosen greedily in this order, keeping the sum of
their bounds below $\tau$. With
our $\tau$, this choice makes the search 6--8\% faster than taking the
smallest bounds, while with $\tau=\theta$ it gains at most 4\%, because the
smaller threshold keeps the candidates few.

Choosing must also cost little. Bounding and sorting every list of a visited
block costs about as much as reading a few postings per list, and many of
these lists are never read. Once per query, the lists are therefore sorted by their bound per posting over the whole list, and the pass of Section~\ref{sec:bmms-order} that computes the bounds of the
sub-blocks also sums, for each sub-block, the bounds of three groups: the 30\%,
50\% and 70\% of the lists with the smallest bound per posting. In a block, the
largest of these groups whose bound stays below
half of $\tau$ is non-essential as a whole, and its lists are bounded and sorted
only if candidates are still left after the other non-essential lists. On CoCondenser this
makes the search 16\% faster at $k=10$, and 2\% at $k=1000$.

\subsection{Completing the few candidates}
\label{sec:bmms-complete}
After the essential lists, very few documents of a block are still
candidates: on CoCondenser, on average 14.4, 21.4 and 42.5 out of 8192 at
$k=10$, 100 and 1000. Adding a non-essential list whole to the accumulator
would read thousands of postings to update a handful of documents. Since the
offsets of a list in a block are sorted, the search can instead
\emph{probe} the list, looking up each candidate with a binary search, or
with a single load in a dense block. For each list, it chooses the cheaper
way. It probes when the number of candidates times the number of steps of a binary
search is well below the number of postings of the list, $|\mathit{cand}|
\cdot (\lfloor\log_2 \mathit{len}\rfloor+1) \cdot 8 < \mathit{len}$, and adds the
list whole otherwise. On CoCondenser, probing makes the search 14--15\%
faster.

To run out of candidates early, the non-essential lists are taken in
decreasing order of their bound per posting. Candidates run out quickly, and the
lists that come after are never read: on CoCondenser, 90\%, 84\% and 70\% of
the non-essential lists at $k=10$, 100 and 1000.

Completing the candidates from the lists themselves is also why \abcbmms
needs no second copy of the postings. \bmp and \lsp store one, grouped by
block (\bmp) or by document (\lsp's forward index), to compute quickly the
full score of the documents that survive pruning. We tried a forward index
in \abcbmms: it doubles the index and makes the search 1.5 to 1.9 times
slower. It scores a document by reading all its terms, about 120 on average,
while completing list by list reads only the lists still needed. Completing
from the lists needs, however, random access to the postings of a block,
which a compressed version must preserve.

\subsection{A compressed \abcbmms}
\label{sec:bmms-packed}
A compressed \abcbmms needs a format that keeps random access, so that a
probe can jump to the right place in a block of a list. In \abcpacked
(Section~\ref{sec:u16}) the gaps can only be decoded in order from the start
of the block, so our compressed format arranges them differently. It cuts the postings of a block of a list into groups
of 9: the first offset of each group is stored in 16 bits, followed by the 8
gaps to the next offsets, bit-packed with a width chosen for the group. A
probe finds the group of a candidate by a binary search on the first
offsets, and decodes only that group, with the SIMD decoder of \abcpacked.
Storing a width per group comes almost for free. The 8 gaps
of a group cannot add up to more than the distance between its first offset
and the next one, which bounds the width; and since an offset takes only 13
of its 16 bits, the 3 spare bits of each first offset tell how much narrower
than this bound the group actually is. Every 8 groups, the start of the group is also stored, so a probe reaches any group by adding at
most 7 widths. Dense blocks stay in dense format, unlike in \abcpacked, because a probe
reads them with a single load.

The compressed variant, \abcbmmspacked, returns the same results. On
CoCondenser it takes 24\% less space than \abcbmms, 2,349 against
3,109\,MiB, and it is at most 11\% slower: 1,199, 2,120 and
3,776\,\textmu s at $k=10$, 100 and 1000, against 1,078, 1,906 and 3,459.
Against \abcpacked (1,814\,MiB,
reordered), the bounds and the random access cost 29\% more space and make
the search 4.9, 2.9 and 1.7 times faster at $k=10$, 100 and 1000.

\section{Faster Doc-id Reordering}
\label{sec:bp}

Reordering the doc-ids so that documents that often occur in the same posting
lists are close pays off in every part of \abc. On CoCondenser it makes the
doc-ids of \abcpacked about a third smaller, it is what makes dense blocks
appear (Section~\ref{sec:u16}), and it gathers the postings of a term in 32\% fewer blocks, which in
\abcbmms means fewer maxima and tighter bounds (Section~\ref{sec:bmms}).

The standard method is recursive graph bisection
(BP)~\cite{dhulipala2016compressing}. BP splits the documents into two halves
and refines the split with \emph{swap rounds}. In each round it estimates,
for every document, how much the log-gap cost
$\sum_t \sum_i \log_2(d_{i+1}^t - d_i^t)$ would decrease if the document
moved to the other half, and swaps the pairs of documents with the largest
gains. After up to 20 rounds it recurses on both halves, down to nodes of 16
documents. It takes into account only the lists of at least 4096 documents,
and every round reads all their postings. Its fastest public
implementation, the \texttt{rgbisect} crate~\cite{mackenzie2021faster},
takes 1.7 minutes on CoCondenser on 20 threads, plus half a minute to
prepare its input, while the rest of the \abc build takes seconds. We
rewrote BP around three ideas.

\paragraph{The real cost of an order.} Log-gap is the cost of an
ideal gap coder, while \abcpacked pays, for each group of 8 gaps, the width
of the largest one, and a header for each block of a list. Moving documents
inside a window of a few hundred doc-ids changes the gaps but rarely the
widths, so the index barely notices, while log-gap keeps rewarding such
moves. Refining the order inside small windows thus costs full passes over
the data and gains almost nothing. We therefore choose every parameter of
the bisection by the exact size of the index it produces, computed from the
posting lists in a few seconds. With this measure, the recursion stops at
nodes of 256 documents instead of 16, which saves four levels without
enlarging the index. The blocks of \abc suggest a second choice: a node
larger than a block of $B=2^{13}$ documents is split at the multiple of $B$
nearest to its middle. Every block is then one subtree of the recursion, and
its documents come from one group of similar documents.

\paragraph{Sample, then rank.} Swap rounds converge slowly, and the split
itself matters less than it seems. In our experiments, splits that differ in
a large share of their documents give indexes of similar size, as long as
each half gathers documents that share terms. We therefore run the swap
rounds, at most four, only on a \emph{sample} of the node, 5\% of its
documents and at least 512, chosen by a hash of the doc-id so that the
sample does not depend on the number of threads. The two halves of the
sample give each term $t$ a preference for the right half,
$p_t = \Delta(\hat\ell_t, \hat r_t) - \Delta(\hat r_t, \hat\ell_t)$, where
$\hat\ell_t$ and $\hat r_t$ are the numbers of documents of $t$ in the two halves,
scaled to the node, and $\Delta(f, g)$ is the estimated gain of a posting
moving from a half where its term has $f$ documents to one where it has $g$
(the estimate used by \texttt{rgbisect}). One pass scores every document of
the node by $\sum_{t \in d} p_t$, and the node is sorted by this score and
cut in the middle. The documents that clearly belong to one half end up at
its far end, and the undecided ones meet at the cut, so the two children
start from an order in which documents that share terms are already close.
In our experiments, ranking every document in this way orders a node better
than swapping pairs, and a level costs about two passes over the postings
instead of up to 20.

\paragraph{Cheap passes.} The pass that scores the documents is bound not
by memory but by the chain of floating-point additions that sums the gains
of the terms of a document. We compute the gain of each term once per round,
as a 16-bit integer, so the gains of a document are summed in four
independent integer accumulators, and the pass becomes about twice as fast.
Integer sums do not depend on their order either, so the order produced is
the same for any number of threads, and so is the index. The documents are
laid out contiguously, and the layout is rebuilt at every level, so that
every pass reads memory sequentially.

\begin{table}
\caption{Doc-id reordering on CoCondenser, on 20 threads: time of the
reordering only, size of the compressed doc-ids in MiB, and
blocks of lists; the best value per column in bold.}
\label{tab:bp}
\small
\begin{tabular}{lrrr}
\toprule
Order & Time & Size & Blocks of lists \\
\midrule
input order & -- & 1,124.8 & 24.05M \\
\texttt{rgbisect} & 101.2\,s & \textbf{755.3} & 16.30M \\
ours & \textbf{5.2\,s} & 764.5 & \textbf{16.25M} \\
\bottomrule
\end{tabular}
\end{table}

\paragraph{Results.} Our reordering is 19.5 times faster than the bisection
of \texttt{rgbisect} alone (Table~\ref{tab:bp}) and keeps 97.5\% of its
reduction of the doc-ids. The doc-ids are 1.2\% larger, but the blocks of lists are 0.3\% fewer, and
\abcbmms is as fast with either order. In the build of \abc, reordering adds
about 6.7\,s in all: \abcpacked is built in 11.9\,s with reordering and in
5.2\,s without it (Table~\ref{tab:qdrant}).
On CoCondenser, the BP of \lucene takes 154\,s, 30 times ours.
Our approach could make reordering cheap for traditional inverted indexes
over text collections too, which we leave to future work.

\section{Experiments}
\label{sec:exp}

We first compare our four indexes with each other, and then \abcbmms and
\abcbmmspacked with the exact and the approximate indexes for learned sparse
retrieval. We use four datasets, encoded by four learned sparse models
that we chose for their different characteristics. Among the exact indexes, \abcbmms is the
fastest on every dataset and at every $k$, and \abcbmmspacked takes about as
much space as \qdrant, \lucene and the other indexes over posting lists. The approximate indexes are
faster at small $k$, but their cost grows faster with $k$, and at $k=1000$
\abcbmms is faster than all of them. It also takes less space, and has no
parameters to tune. We finally present an experiment that shows another
virtue of \abc: it computes the exact ground truth of a set of queries
27--52$\times$ faster than brute force.

\subsection{Experimental settings}
\label{sec:exp-settings}
\paragraph{Machine and software.} Intel Core Ultra 7 265K (Arrow Lake, 8 P-cores up to
5.5\,GHz and 12 E-cores; 48\,KiB L1d and 3\,MiB L2 per P-core, 30\,MiB
shared L3; AVX2, no AVX-512), 122\,GiB of RAM, Linux 7.0.
Our indexes are written in Rust and compiled with
\texttt{rustc} 1.101.0-nightly (2026-10-15), with link-time optimization,
one codegen unit and \texttt{-C target-cpu=native}; the code will be released upon acceptance
of the paper.

\paragraph{Datasets.} All datasets encode the 8,841,823 passages and the 6,980
dev queries of MS MARCO v1 passage ranking~\cite{nguyen2016msmarco}, so that
they differ only in the learned sparse model. The models are chosen to differ
in the length of queries and passages and in the number of dimensions
(Table~\ref{tab:datasets}):
\begin{itemize}
  \item CoCondenser, the SPLADE model built on
    CoCondenser~\cite{gao2022cocondenser,formal2022distillation}, is the
    historical reference: most efficiency studies of learned sparse retrieval,
    among them those of \seismic, \bmp, \spidx and \lsp, report on it, so our
    numbers can be compared with theirs.
  \item SparseUp~\cite{linkup2026sparseup}, released in September 2026, is the
    most recent model: a SPLADE-style encoder. Its queries are about as long as CoCondenser's, on longer passages, and
    we use it to check our results on a current model.
  \item OS-gte~\cite{shen2025l0}, a recent model made
    for the sparse search of OpenSearch
    (\texttt{opensearch-\allowbreak neural-\allowbreak sparse-\allowbreak encoding-\allowbreak doc-\allowbreak v3-\allowbreak gte}),
    represents the scenario in which the
    model does not expand the queries: it is inference-free, it encodes only
    the passages, while a query is its tokens weighted by a fixed IDF table,
    with 6.9 non-zeros on average. We prefer it to Efficient
    SPLADE~\cite{lassance2022efficiency}, the usual model for this scenario,
    because it is recent; on Efficient SPLADE we observed similar results.
  \item SPLARE~\cite{formal2026splare} replaces the vocabulary projection with
    the latent space of a sparse autoencoder: 123,627 dimensions instead of about 30,000, and the most non-zeros
    both per passage (284.6) and per query (106.9). It stresses the size of the
    term dictionary and of the per-term structures, and the cost of long
    queries, where every term adds a list to traverse and loosens the bounds.
    The encoded collection is not publicly available: we obtained it by
    contacting the authors privately.
\end{itemize}
Table~\ref{tab:datasets} reports statistics on the datasets: the number of
dimensions, the number of non-zeros in total, per passage and per query, and
the recall of exact search with our 8-bit document weights against the exact
\texttt{f32} top-$k$. It is at least 0.992 on every dataset and at every $k$,
so our 8-bit weights lose less than 1\% of the true top-$k$.

\begin{table}
\caption{The four datasets. Dimensions are the coordinates that occur; recall@$k$
is that of exact search with 8-bit document weights against the exact
\texttt{f32} top-$k$.}
\label{tab:datasets}
\small\setlength{\tabcolsep}{3pt}
\begin{tabular}{lrrrrc}
\toprule
 & & Non-zeros & \multicolumn{2}{c}{Non-zeros per} & Recall (u8) \\
Dataset & Dims & ($\times10^9$) & passage & query & $k=10/100/1000$ \\
\midrule
CoCondenser & 28,679 & 1.06 & 120.0 & 43.9 & .9925/.9953/.9964 \\
OS-gte & 29,402 & 1.60 & 180.5 & 6.9 & .9922/.9943/.9950 \\
SparseUp & 31,227 & 1.68 & 189.5 & 47.4 & .9969/.9974/.9979 \\
SPLARE & 123,627 & 2.52 & 284.6 & 106.9 & .9986/.9989/.9990 \\
\bottomrule
\end{tabular}
\end{table}

\paragraph{Indexes.} We compare our indexes with seven exact indexes for
learned sparse retrieval that have public code, and with two approximate ones.
We list the indexes grouped by kind.
\begin{itemize}
  \item \emph{Our indexes}: the exhaustive \abc and its compressed variant
    \abcpacked, and \abcbmms and its compressed variant \abcbmmspacked, all
    with 8-bit document weights and doc-id reordering.
  \item \emph{Exact indexes over blocks of documents}, all run with no
    approximation. \bmp~\cite{mallia2024bmp} scores blocks of documents through per-block
    inverted lists in decreasing order of their upper bound; we use blocks of
    16 documents (the fastest at maximum recall in a grid search) and
    $\alpha=\beta=1$. \spidx~\cite{carlson2025sp} adds superblocks to \bmp:
    superblocks of 512 and blocks of 8 documents, and
    $\mu=\eta=\beta=\gamma=1$. \lsp~\cite{carlson2026lsp} bounds superblocks
    and blocks with compressed maxima and scores through a
    \seismic-style forward index: superblocks of 128 and blocks of 8
    documents, $\mu=\eta=\beta=\gamma=1$ as for \spidx, and 8-bit block maxima, because the
    4-bit maxima of its paper are not upper bounds of the scores
    (Section~\ref{sec:exp-compare}).
  \item \emph{Exact indexes over posting lists.} The index
    \rise~\cite{savino2026rise} re-implements in Rust
    DS2I~\cite{ottaviano2014partitioned}, of which \pisa~\cite{mallia2019pisa}
    is an extension, and it is about as fast as \pisa. It processes queries document at a time,
    with dynamic pruning over compressed lists with block
    metadata, and weights scaled by 1000 and rounded down; we use the fastest
    of its MaxScore, Block-Max MaxScore and exhaustive evaluation. We also ran
    \pisa with the fastest of its algorithms: it is at most 15\% faster than
    \rise (on OS-gte), so we report only \rise. \lucene 10.5.2~\cite{lucene_maxscorebulkscorer}
    has its index merged into one segment and reordered by its own BP, with
    8-bit document weights (largest 255) and a
    similarity that multiplies them by the query weight, so that a
    disjunction of term queries computes the exact inner product. We run both
    its Block-Max MaxScore scorer, which \lucene uses when only the top $k$
    is needed, and its exhaustive evaluation, which scores every posting
    term at a time in windows of 4096 doc-ids, and as for \rise we report
    the faster of the two at each $k$. The sparse index of
    \qdrant~\cite{qdrant2025sparse}, which scores every posting term at a time
    in windows of doc-ids (Section~\ref{sec:qdrant}), is called in process
    through its \texttt{sparse} crate (v1.20.2), with its compressed index and
    8-bit weights.
    \ioqp~\cite{mackenzie2022ioqp} is an impact-ordered score-at-a-time
    evaluation, run exhaustively ($\rho=1$).
  \item \emph{Approximate indexes.}
    \seismic~\cite{bruch2024seismic,nardini2027compact} 0.6.0 runs with a
    memory budget of at most $1.5\times$ and $2\times$ the size of its forward
    index, the budgets used to evaluate it at
    scale~\cite{bruch2025scalability}. At $k=10$ on CoCondenser we use the
    best configurations released with \seismic for recall@10 of 99\%.
    Elsewhere no released configuration exists, and we run its grid-search scripts
    around those configurations. For each $k$ and budget we keep the
    fastest configuration that reaches 0.99, so \seismic uses a different
    index for each $k$. The HNSW graph of \kannolo~\cite{delfino2025kannolo} is compressed
    losslessly with node reordering and
    StreamVByte~\cite{benocci2026lossless}, with \texttt{f16} values, $M=32$
    and $\mathit{ef}_{c}=2000$, the setting of its paper; at search time
    $\mathit{ef}_{s}$ goes from 10 to 32,000, with or without its
    distance-adaptive early termination ($\lambda$ from 0.01 to 0.15).
\end{itemize}

\paragraph{Measurements.} We run one query at a time on one thread pinned to
a P-core, and one benchmark at a time on an otherwise idle machine. We report
the mean latency over all the queries, without the time to load the index.
Every index written in Rust is built with \texttt{-C target-cpu=native}, and
\lucene runs on OpenJDK 21 with the vector module. \lsp, \ioqp and \lucene
are timed on a second pass over the queries, after a warm-up pass, and the other indexes, ours included, on the first pass. Recall is recall@$k$ against the exact
\texttt{f32} top-$k$.

\subsection{Our indexes}
\label{sec:exp-bmms}
Table~\ref{tab:abc} compares our four indexes on the four datasets. They all
use the same 8-bit weights and return the same results, so their recall is that of Table~\ref{tab:datasets}, up to ties.

\begin{table}
\caption{Our indexes, with 8-bit weights and reordered doc-ids: latency in
\textmu s/query on one thread at $k=10/100/1000$, and index size in GiB;
recall as in Table~\ref{tab:datasets}; the fastest per dataset and $k$ in bold.}
\label{tab:abc}
\small\setlength{\tabcolsep}{4pt}
\begin{tabular}{llrrrr}
\toprule
Dataset & Index & $k=10$ & $k=100$ & $k=1000$ & Size \\
\midrule
\multirow{4}{*}{CoCondenser} & \abc & 4,099 & 4,140 & 4,337 & 2.9 \\
 & \abcpacked & 5,850 & 6,050 & 6,233 & 1.8 \\
 & \abcbmms & \textbf{1,078} & \textbf{1,906} & \textbf{3,459} & 3.0 \\
 & \abcbmmspacked & 1,199 & 2,120 & 3,776 & 2.3 \\
\midrule
\multirow{4}{*}{OS-gte} & \abc & 1,073 & 1,099 & 1,287 & 4.4 \\
 & \abcpacked & 1,836 & 1,756 & 2,206 & 2.6 \\
 & \abcbmms & \textbf{264} & \textbf{535} & \textbf{1,032} & 4.5 \\
 & \abcbmmspacked & 315 & 636 & 1,202 & 3.3 \\
\midrule
\multirow{4}{*}{SparseUp} & \abc & 5,250 & 5,279 & 5,512 & 4.6 \\
 & \abcpacked & 8,859 & 8,811 & 8,858 & 2.9 \\
 & \abcbmms & \textbf{1,824} & \textbf{2,841} & \textbf{4,389} & 4.8 \\
 & \abcbmmspacked & 2,109 & 3,367 & 4,728 & 3.6 \\
\midrule
\multirow{4}{*}{SPLARE} & \abc & 5,844 & 5,869 & 6,079 & 7.2 \\
 & \abcpacked & 7,265 & 7,837 & 8,249 & 4.6 \\
 & \abcbmms & \textbf{1,726} & \textbf{3,221} & \textbf{5,363} & 7.6 \\
 & \abcbmmspacked & 1,919 & 3,659 & 6,117 & 5.7 \\
\bottomrule
\end{tabular}
\end{table}

\paragraph{Pruning.} The pruned \abcbmms is faster than \abc on every
dataset and at every $k$: 2.9--4.1$\times$ at $k=10$, 1.8--2.2$\times$ at
$k=100$ and 1.1--1.3$\times$ at $k=1000$, for 3.4--5.6\% more space (the
sub-block maxima). The gain shrinks as $k$ grows, because $\theta$ is lower
and fewer documents can be dropped. The exhaustive \abc, instead, scores
every posting of the query lists and costs about the same at every $k$. At
$k=10$, the gain is largest on OS-gte, the dataset with the shortest queries, 7
non-zeros on average.

\paragraph{Compression.} The compressed \abcpacked is 24--71\% slower than
\abc, the cost of decoding the doc-id gaps, and takes 37--40\% less space.
With pruning, compression costs much less: \abcbmmspacked is at most 19\%
slower than \abcbmms, because Block-Max MaxScore decodes far fewer postings,
and takes about a quarter less space.

\subsection{Comparison with the exact indexes}
\label{sec:exp-compare}
Since compression costs little with pruning, Table~\ref{tab:compare} reports
both \abcbmms and \abcbmmspacked, with the other indexes, on the four
datasets at $k=10$, 100 and 1000. We start from the exact indexes, after
checking which of them are exact.

\begin{table*}
\caption{Exact and approximate indexes: latency in \textmu s/query on one
thread, index size in GiB, build time on 20 threads, and recall@$k$ against the exact \texttt{f32}
top-$k$; the fastest per dataset and $k$ in bold. The row \abcbmmspacked
(\texttt{f32}) uses the original 32-bit weights; the rows of \seismic have a
memory budget of $1.5\times$ and $2\times$ the size of its forward index, and report the
size and build time of its largest index.}
\label{tab:compare}
\small\setlength{\tabcolsep}{4pt}\renewcommand{\arraystretch}{0.93}
\begin{tabular}{llrrrrrl}
\toprule
 & Index & $k=10$ & $k=100$ & $k=1000$ & Size & Build & Recall@10/100/1000 \\
\midrule
\multirow{13}{*}{\rotatebox[origin=c]{90}{CoCondenser}} & \abcbmms & 1,078 & 1,906 & \textbf{3,459} & 3.0 & 13\,s & .9925/.9953/.9964 \\
 & \abcbmmspacked & 1,199 & 2,120 & 3,776 & 2.3 & 15\,s & .9925/.9953/.9964 \\
 & \abcbmmspacked (\texttt{f32}) & 1,476 & 2,805 & 5,040 & 5.1 & 17\,s & 1/1/1 \\
\cmidrule{2-8}
 & \seismic ($1.5\times$) & 299 & 1,673 & 9,321 & 5.8 & 102\,s & .9902/.9901/.9910 \\
 & \seismic ($2\times$) & \textbf{258} & 1,167 & 5,760 & 7.8 & 53\,s & .9910/.9902/.9900 \\
 & \kannolo (HNSW) & 428 & \textbf{1,034} & 5,565 & 4.9 & 12.0\,h & .9904/.9907/.9902 \\
\cmidrule{2-8}
 & \bmp & 3,136 & 6,060 & 19,541 & 20.9 & 150\,s & .9880/.9912/.9914 \\
 & \spidx & 2,383 & 3,651 & 10,597 & 38.1 & 244\,s & .9682/.9698/.9655 \\
 & \lsp & 5,566 & 6,862 & 12,277 & 12.4 & 67\,s & .9880/.9912/.9915 \\
\cmidrule{2-8}
 & \ioqp & 59,812 & 59,296 & 59,910 & 2.1 & 6\,s & .9879/.9912/.9914 \\
 & \qdrant & 30,531 & 30,744 & 30,692 & 2.2 & 36\,s & .9925/.9953/.9963 \\
 & \rise & 59,599 & 69,839 & 84,984 & 3.5 & 39\,s & .9989/.9993/.9993 \\
 & \lucene & 59,386 & 59,022 & 60,934 & 1.9 & 523\,s & .9898/.9940/.9953 \\
\midrule
\multirow{13}{*}{\rotatebox[origin=c]{90}{OS-gte}} & \abcbmms & 264 & \textbf{535} & \textbf{1,032} & 4.5 & 19\,s & .9922/.9943/.9950 \\
 & \abcbmmspacked & 315 & 636 & 1,202 & 3.3 & 22\,s & .9922/.9943/.9950 \\
 & \abcbmmspacked (\texttt{f32}) & 358 & 737 & 1,475 & 7.5 & 25\,s & 1/1/1 \\
\cmidrule{2-8}
 & \seismic ($1.5\times$) & 327 & 5,637 & 12,650 & 8.5 & 102\,s & .9901/.9917/.9907 \\
 & \seismic ($2\times$) & \textbf{240} & 2,059 & 7,666 & 12.0 & 232\,s & .9901/.9901/.9902 \\
 & \kannolo (HNSW) & 4,983 & 19,824 & 56,114 & 6.6 & 9.3\,h & .9812/.9851/.9849 \\
\cmidrule{2-8}
 & \bmp & 778 & 1,919 & 8,131 & 23.8 & 230\,s & .9875/.9897/.9901 \\
 & \spidx & 534 & 1,058 & 4,175 & 42.0 & 446\,s & .9777/.9769/.9743 \\
 & \lsp & 716 & 1,390 & 4,896 & 15.3 & 114\,s & .9875/.9897/.9901 \\
\cmidrule{2-8}
 & \ioqp & 21,941 & 22,096 & 22,601 & 3.2 & 8\,s & .9875/.9897/.9901 \\
 & \qdrant & 22,060 & 22,096 & 22,196 & 3.4 & 54\,s & .9917/.9944/.9950 \\
 & \rise & 7,037 & 10,243 & 14,483 & 4.8 & 82\,s & .9981/.9984/.9987 \\
 & \lucene & 13,400 & 16,833 & 21,075 & 2.8 & 991\,s & .9899/.9927/.9936 \\
\midrule
\multirow{13}{*}{\rotatebox[origin=c]{90}{SparseUp}} & \abcbmms & 1,824 & 2,841 & \textbf{4,389} & 4.8 & 21\,s & .9969/.9974/.9979 \\
 & \abcbmmspacked & 2,109 & 3,367 & 4,728 & 3.6 & 26\,s & .9969/.9974/.9979 \\
 & \abcbmmspacked (\texttt{f32}) & 2,604 & 4,094 & 6,405 & 8.1 & 27\,s & 1/1/1 \\
\cmidrule{2-8}
 & \seismic ($1.5\times$) & 4,978 & 5,403 & 46,864 & 9.3 & 165\,s & .9893/.9908/.9854 \\
 & \seismic ($2\times$) & \textbf{615} & 3,042 & 19,864 & 12.3 & 156\,s & .9902/.9911/.9900 \\
 & \kannolo (HNSW) & 1,180 & \textbf{2,233} & 9,689 & 7.2 & 16.7\,h & .9903/.9902/.9901 \\
\cmidrule{2-8}
 & \bmp & 7,041 & 23,422 & 78,654 & 26.8 & 269\,s & .9949/.9953/.9960 \\
 & \spidx & 5,512 & 11,510 & 40,131 & 46.0 & 460\,s & .2181/.3305/.4033 \\
 & \lsp & 16,478 & 19,261 & 31,315 & 21.8 & 171\,s & .7489/.9477/.9948 \\
\cmidrule{2-8}
 & \ioqp & 110,937 & 111,754 & 111,070 & 3.5 & 8\,s & .9949/.9953/.9960 \\
 & \qdrant & 28,569 & 28,729 & 28,877 & 3.5 & 57\,s & .9968/.9974/.9979 \\
 & \rise & 96,927 & 113,702 & 136,603 & 6.0 & 59\,s & .9998/.9997/.9997 \\
 & \lucene & 81,906 & 86,422 & 84,664 & 3.2 & 1,018\,s & .9963/.9971/.9977 \\
\midrule
\multirow{13}{*}{\rotatebox[origin=c]{90}{SPLARE}} & \abcbmms & 1,726 & 3,221 & \textbf{5,363} & 7.6 & 32\,s & .9986/.9989/.9990 \\
 & \abcbmmspacked & 1,919 & 3,659 & 6,117 & 5.7 & 38\,s & .9986/.9989/.9990 \\
 & \abcbmmspacked (\texttt{f32}) & 2,198 & 4,341 & 7,454 & 12.5 & 41\,s & 1/1/1 \\
\cmidrule{2-8}
 & \seismic ($1.5\times$) & 749 & 2,597 & 16,195 & 21.2 & 245\,s & .9914/.9902/.9909 \\
 & \seismic ($2\times$) & 747 & 2,128 & 10,565 & 27.4 & 304\,s & .9914/.9914/.9904 \\
 & \kannolo (HNSW) & \textbf{399} & \textbf{1,330} & 8,486 & 15.0 & 27.2\,h & .9904/.9901/.9911 \\
\cmidrule{2-8}
 & \bmp & \multicolumn{3}{l}{build $>$ 115\,GiB of RAM} & -- & -- & -- \\
 & \spidx & \multicolumn{3}{l}{build $>$ 115\,GiB of RAM} & -- & -- & -- \\
 & \lsp & 35,111 & 51,302 & 95,619 & 40.9 & 565\,s & .7197/.7420/.6813 \\
\cmidrule{2-8}
 & \ioqp & 77,519 & 76,869 & 77,038 & 5.4 & 12\,s & .9978/.9983/.9985 \\
 & \qdrant & 34,650 & 34,753 & 35,117 & 5.6 & 92\,s & .9986/.9989/.9990 \\
 & \rise & 106,681 & 127,794 & 153,081 & 9.3 & 156\,s & .9998/.9999/.9999 \\
 & \lucene & 68,715 & 69,094 & 72,006 & 5.2 & 1,812\,s & .9980/.9985/.9987 \\
\bottomrule

\end{tabular}

\end{table*}

\paragraph{Some exact indexes are not so exact.} \bmp, \spidx, \lsp and \ioqp
are designed to be exact, but they lose recall for three reasons, all in the
implementation rather than in the algorithm. Query weights are quantized to few levels (the largest to
32 in \bmp, \spidx and \ioqp, to 128 in \lsp): on CoCondenser and OS-gte,
where nothing overflows, this alone limits their recall, at any $k$, to
0.966--0.978 for \bmp and \spidx and to 0.985--0.989 for \lsp. Block maxima
on 4 bits, as in the paper of \lsp, are not upper bounds of the scores: at
$k=10$, the results of between 670 (CoCondenser) and 3,005 (OS-gte) of the 6,980 queries
differ from those of the same engine without pruning. Scores, bounds or thresholds on
16 bits overflow on the long queries of recent models: on SparseUp the
original \bmp reaches a recall@10 of 0.38, \spidx 0.22 and \lsp 0.75; on
SPLARE, whose 123,627 dimensions do not fit in 16-bit term ids, \lsp stays
below 0.75. We removed these causes in \bmp and \ioqp, and the first two in
\lsp, and we report these versions. We left the overflow in \lsp and every
cause in \spidx: removing them would only widen scores and bounds, which
would not make either faster, and both are slower than \abcbmms at every
$k$. We report them with their own recall, which for \spidx stays below that
of exact search on every dataset, and for \lsp on SparseUp and SPLARE, where
it overflows; \spidx also misses documents in 10 queries at $k=1000$. With
our fixes,
\bmp and \ioqp, like \rise, \qdrant and \lucene, are exact: their recall
differs from that of \abcbmms only because each quantizes the weights in its
own way.

\paragraph{Time.} Our \abcbmms is the fastest exact index on every dataset
and at every $k$. With the original 32-bit weights, which give the true
top-$k$, \abcbmmspacked is also faster than every other exact index with
8-bit weights. Against the indexes over blocks of documents, \bmp, \spidx
and \lsp, the margin widens at large $k$, because the number of blocks they
score grows with $k$: against the fastest of them, where all three can be built, it is 1.9--4.1$\times$
at $k=10$ and 100, and 3.1--7.1$\times$ at $k=1000$.
Against the indexes over posting lists, \qdrant, \lucene, \rise and \ioqp,
the margin is 6.5--84$\times$, and it shrinks with $k$, because most of them
score every posting, at a cost that does not depend on $k$. On SPLARE, where
\bmp and \spidx cannot be built, the fastest of the other exact indexes is \qdrant,
6.5--20$\times$ slower than \abcbmms.

\paragraph{Pruning in the other indexes.} In the other indexes, pruning does
not always pay off. \lucene's Block-Max MaxScore is slower than its own
exhaustive evaluation, by up to 2.7$\times$, on all datasets and at every
$k$, except on OS-gte at $k\le100$. In \rise, WAND and Block-Max WAND,
which we do not report, are slower than the exhaustive evaluation on CoCondenser, as reported for
\pisa~\cite{mackenzie2021wacky}, while Block-Max MaxScore is
2.1--4.6$\times$ faster.

\paragraph{Space and build.} The indexes over blocks of documents are
3--13$\times$ larger than \abcbmms (\bmp 5--7$\times$, \lsp 3--5$\times$ and
\spidx 9--13$\times$), and on SPLARE the builds of \bmp and \spidx exceed
115\,GiB of RAM. The
compressed \abcbmmspacked takes about as much space as the indexes over
posting lists: as much as \qdrant and \ioqp, less than \rise, and
10--21\% more than \lucene, whose index is also reordered. Our indexes are
built in 13--41\,s on 20 threads, doc-id reordering included. \ioqp takes
6--12\,s and the indexes over blocks of documents minutes, all plus the
reordering of their input, \qdrant and \rise 36--156\,s, and \lucene up to half an hour,
most of it in its BP.

\subsection{Comparison with the approximate indexes}
\label{sec:exp-approx}
We now compare \abcbmms with the approximate indexes of
Table~\ref{tab:compare}, \seismic and \kannolo.

\paragraph{Time.} We run \seismic and \kannolo at the fastest setting whose
recall@$k$ reaches 0.99, just below that of exact search with 8-bit weights
(Table~\ref{tab:datasets}). On OS-gte, \kannolo does not reach 0.99 at any $k$ (0.981--0.985), nor does
\seismic with a budget of $1.5\times$ on SparseUp at $k=10$ and 1000. At
$k=10$, \seismic with a budget of $2\times$ is 1.1--4.2$\times$
faster than \abcbmms, and \kannolo 1.5--4.3$\times$ faster, except on OS-gte,
where it is 19$\times$ slower. Their cost, however, grows faster with $k$,
because to keep the recall at 0.99 they must examine more documents:
\kannolo with a larger $\mathit{ef}_s$, \seismic with more query terms or
less pruning of its blocks. At $k=100$ they are at most 2.4$\times$ faster,
and \abcbmms is the fastest on OS-gte and faster than \seismic on SparseUp.
At $k=1000$ \abcbmms is the fastest index everywhere, 1.7--7.4$\times$ faster
than \seismic with a budget of $2\times$ and 1.6--2.2$\times$ faster than
\kannolo on the other three datasets.

\paragraph{Space and build.} Our \abcbmms takes 1.5--3.6$\times$ less space
than \seismic and \kannolo. Our index is
built in seconds, against minutes for \seismic and 9--27 hours for
\kannolo. It has no parameters to tune, and one index serves every $k$, whereas
\seismic needs a grid search and a different index for every $k$, and
\kannolo a different value of $\mathit{ef}_s$ for every $k$.

\subsection{Exact ground truth computation}
\label{sec:gt}
The recall of an approximate index is measured against
the exact top-$k$ of a set of queries, the ground truth, which also serves to
measure the effectiveness of a model. The ground truth is usually computed
by exhaustive search, with a multithreaded brute-force scan such as that of
vectorium~\cite{vectorium}, the library under \seismic and \kannolo. The
tool \texttt{abc-\allowbreak groundtruth} computes it on all cores with an
\texttt{f32} \abc index, which gives exact scores. The index is not
reordered, because at this number of queries reordering costs more than the
search time it saves. On the four datasets (6,980 queries, 20 threads),
the tool takes 15--30\,s at $k=1000$, reading the input and building the
index included, and 11--25\,s at $k=10$ with \abcbmms. The brute force of
vectorium takes 9--25 minutes, so our tool is 27--52$\times$ faster at
$k=1000$.

\section{Conclusions}
Standard IR techniques, combined and engineered for modern CPUs, give a
fast exact index with no parameters for learned sparse retrieval. In \abc,
term-at-a-time evaluation runs over blocks of doc-ids whose accumulators fit
in the L1 cache, so that scoring a posting is a multiply-add into a slot
already in cache. Block-Max MaxScore inside these blocks makes \abcbmms
faster than \abc at every $k$, and the fastest exact index: an order of
magnitude faster than \qdrant and \lucene, and several times smaller than
the other block-based indexes. Approximate indexes win at small $k$ and
\abcbmms at large $k$, with a smaller index built in seconds.

Our blocks and pruning do not depend on learned weights, so they also apply
to classic term weights, such as BM25, which are outside the scope of this
paper. Future work includes reordering other indexes with the graph
bisection that reorders \abc, evaluating \abcbmms on larger collections such
as MS MARCO v2, and making the index updatable, which is simpler because a
full rebuild takes seconds and no parameter needs to be tuned.

\section*{Use of Generative AI}
Claude Opus 5.5 helped the author write and debug the code, automate the
experiments, and write and revise the text. The ideas and their combination
are the author's, who reviewed all of its output.

\bibliographystyle{ACM-Reference-Format}
\bibliography{references}

\end{document}

%% file: fig_u16.tex
\begin{tikzpicture}[
    x=1cm, y=1cm, font=\footnotesize,
    cell/.style={draw, minimum height=0.48cm, inner sep=0pt, anchor=west},
    lab/.style={anchor=east, align=right},
    note/.style={anchor=west, align=left, text=black!70, font=\scriptsize},
  ]
  \def\bw{2.55}  %
  \def\xo{2.4}   %
  \node[lab] at (\xo-0.15, 0.25) {doc-ids of $t$};
  \foreach \blk in {0,...,5} {
    \draw[fill=black!3] (\xo+\blk*\bw, 0) rectangle ++(\bw, 0.5);
    \node[font=\scriptsize] at (\xo+\blk*\bw+\bw/2, -0.25) {block \blk};
  }
  \foreach \blk/\off in {0/147, 0/1243, 0/1435, 0/5800, 2/450, 2/2010,
                     5/310, 5/1210, 5/2280} {
    \draw[thick] (\xo+\blk*\bw+\off/8192*\bw, 0.05) -- ++(0, 0.4);
  }
  \foreach \idx in {0,...,40} {
    \pgfmathsetmacro{\rr}{mod(\idx*37,41)/41}
    \ifdim\rr pt>0.25pt
      \draw ({\xo+3*\bw+0.04+\idx/41*(\bw-0.08)}, 0.05) -- ++(0, 0.4);
    \fi
  }
  \node[lab] at (\xo-0.15, -0.8) {used-block bitvector};
  \foreach \blk/\val in {0/1, 1/0, 2/1, 3/1, 4/0, 5/1} {
    \node[cell, minimum width=0.5cm] (bv\blk) at (\xo+\blk*\bw+\bw/2-0.25, -0.8) {\val};
  }

  \def\yr{-2.0}
  \node[lab] at (\xo-0.15, \yr) {block ranges};
  \foreach \jj/\val/\txt in {0/0/0, 1/4/2, 2/6/3, 3/6/5, 4/9/end} {
    \node[cell, minimum width=0.8cm] (so\jj) at (\xo+\jj*0.8, \yr) {\val};
    \node[font=\scriptsize, text=black!60] at (\xo+\jj*0.8+0.4, \yr+0.42) {\txt};
  }
  \node[note] at (\xo+4*0.8+1.0, \yr+0.03)
    {one start per used block (and an end marker);\\
     block 3 is dense: its range $[6,6)$ is empty};

  \def\yo{-3.5}
  \def\cw{0.95}
  \node[lab] at (\xo-0.15, \yo) {offsets (\texttt{u16})};
  \node[lab] at (\xo-0.15, \yo-0.6) {weights (\texttt{u8})};
  \foreach \idx/\off/\wt in {0/147/83, 1/1243/12, 2/1435/201, 3/5800/40,
                        4/450/7, 5/2010/155, 6/310/66, 7/1210/9, 8/2280/120} {
    \node[cell, minimum width=\cw cm] (of\idx) at (\xo+\idx*\cw, \yo) {\off};
    \node[cell, minimum width=\cw cm] at (\xo+\idx*\cw, \yo-0.6) {\wt};
  }
  \foreach \idx in {0,...,8} {
    \node[font=\scriptsize, text=black!50] at (\xo+\idx*\cw+\cw/2, \yo+0.4) {\idx};
  }
  \foreach \aa/\blk/\txt in {0/3/block 0, 4/5/block 2, 6/8/block 5} {
    \draw[decorate, decoration={brace, mirror, amplitude=3pt}]
      (\xo+\aa*\cw+0.05, \yo-0.95) -- (\xo+\blk*\cw+\cw-0.05, \yo-0.95)
      node[midway, below=2pt, font=\scriptsize] {\txt};
  }
  \foreach \jj/\idx in {0/0, 1/4, 3/6} {
    \draw[-{Stealth[length=4pt]}, black!60] (so\jj.south) to[out=-90, in=90] (of\idx.north west);
  }
  \node[note] at (\xo+9*\cw+0.15, \yo-0.3)
    {the offset is the\\accumulator index:\\no decoding};

  \def\yc{-5.6}
  \node[lab] at (\xo-0.15, \yc) {dense block 3};
  \foreach \idx/\val in {0/0, 1/31, 2/0, 3/64, 4/0, 5/18, 6/\dots, 7/97, 8/0} {
    \node[cell, minimum width=0.7cm] at (\xo+\idx*0.7, \yc) {\val};
  }
  \foreach \idx/\val in {0/0, 1/1, 2/2, 3/3, 4/4, 5/5, 7/8190, 8/8191} {
    \node[font=\scriptsize, text=black!50] at (\xo+\idx*0.7+0.35, \yc+0.4) {\val};
  }
  \node[note] at (\xo+9*0.7+0.2, \yc)
    {8192 weights, one per doc-id of the block, 0 if absent};
\end{tikzpicture}

%% file: fig_bmms.tex
\begin{tikzpicture}[
    x=1cm, y=1cm, font=\footnotesize,
    cell/.style={draw, minimum height=0.48cm, inner sep=0pt, anchor=west},
    lab/.style={anchor=east, align=right},
    note/.style={anchor=west, align=left, text=black!70, font=\scriptsize},
  ]
  \def\xo{3.2}   %
  \def\sw{1.2}   %
  \node[lab] at (\xo-0.15, 0.25) {block 0 of $t$};
  \draw[fill=black!3] (\xo, 0) rectangle ++(8*\sw, 0.5);
  \foreach \ss in {1,...,7} {
    \draw[densely dashed, black!50] (\xo+\ss*\sw, 0) -- ++(0, 0.5);
  }
  \foreach \ss in {0,...,7} {
    \node[font=\scriptsize, text=black!60] at (\xo+\ss*\sw+\sw/2, -0.22) {sub-block \ss};
  }
  \foreach \off/\wt/\anc in {147/83/center, 1243/12/east, 1435/201/west, 5800/40/center} {
    \draw[thick] (\xo+\off/8192*8*\sw, 0.05) -- ++(0, 0.4);
    \node[font=\scriptsize, anchor=\anc, inner sep=1pt] at (\xo+\off/8192*8*\sw, 0.72) {\wt};
  }
  \def\ym{-1.0}
  \node[lab] at (\xo-0.15, \ym) {maxima of ($t$, block 0)};
  \foreach \ss/\val in {0/83, 1/201, 2/0, 3/0, 4/0, 5/40, 6/0, 7/0} {
    \node[cell, minimum width=\sw cm] at (\xo+\ss*\sw, \ym) {\val};
  }
  \node[note] at (\xo+8*\sw+0.15, \ym) {one byte per\\sub-block};

  \def\gw{2.6}   %
  \def\mw{0.325} %
  \def\ys{-2.5}
  \def\yx{-3.1}
  \node[lab] at (\xo-0.15, \ys) {block ranges};
  \node[lab] at (\xo-0.15, \yx) {sub-block maxima};
  \foreach \jj/\val/\blk in {0/0/0, 1/4/2, 2/6/3, 3/6/5} {
    \node[cell, minimum width=\gw cm] at (\xo+\jj*\gw, \ys) {\val};
    \node[font=\scriptsize, text=black!60] at (\xo+\jj*\gw+\gw/2, \ys+0.42) {block \blk};
  }
  \node[cell, minimum width=0.6cm] at (\xo+4*\gw, \ys) {9};
  \node[font=\scriptsize, text=black!60] at (\xo+4*\gw+0.3, \ys+0.42) {end};
  \foreach \jj/\vals in {0/{83,201,0,0,0,40,0,0}, 1/{7,155,0,0,0,0,0,0},
                         2/{230,198,215,240,221,187,205,233}, 3/{66,9,120,0,0,0,0,0}} {
    \foreach \val [count=\ss from 0] in \vals {
      \node[cell, minimum width=\mw cm, font=\tiny] at (\xo+\jj*\gw+\ss*\mw, \yx) {\val};
    }
  }
  \node[note] at (\xo+4*\gw+0.75, \ys-0.3)
    {8 bytes per used block,\\read with one load};
\end{tikzpicture}